\documentclass[prl, twocolumn, superscriptaddress,floatfix, nofootinbib]{revtex4-1}

\usepackage{url}
\usepackage{xspace}
\usepackage{dsfont}
\usepackage{amssymb}
\usepackage{amsmath}
\usepackage{graphicx}
\usepackage[small]{subfigure}
\usepackage[colorlinks=true,citecolor=blue]{hyperref}
\usepackage{multirow}
\usepackage{hhline}
\usepackage{color}
\definecolor{darkred}{rgb}{1.0,0.1,0.1}
\definecolor{darkgreen}{rgb}{0.1,0.7,0.1}
\definecolor{darkblue}{rgb}{0.1,0.1,1.0}

\usepackage{amsmath}
\DeclareMathOperator*{\argmax}{argmax}

\DeclareRobustCommand{\Tab}[1]{Table~\ref{tab:#1}}

\DeclareRobustCommand{\Eq}[1]{Eq.~(\ref{eq:#1})}

\newcommand{\eqn}[1]{\begin{align}#1\end{align}}

\def\alphas{\texttt{TimeShower:alphaSvalue}}
\def\lund{\texttt{StringZ:aLund}}
\def\strange{\texttt{StringFlav:probStoUD}}

\def\ndf{\text{ndf}}

\newcommand{\Pythia}{{\sc Pythia}\xspace}
\newcommand{\Herwig}{{\sc Herwig}\xspace}

\newcommand{\DCTR}{{\sc Dctr}\xspace}

\begin{document}

\title{Neural Networks for Full Phase-space Reweighting and Parameter Tuning}

\author{Anders Andreassen}
\email{andersja@berkeley.edu}
\affiliation{Department of Physics, University of California, Berkeley, CA 94720, USA}
\affiliation{Physics Division, Lawrence Berkeley National Laboratory, Berkeley, CA 94720, USA}

\author{Benjamin Nachman}
\email{bpnachman@lbl.gov}
\affiliation{Physics Division, Lawrence Berkeley National Laboratory, Berkeley, CA 94720, USA}

\begin{abstract}
Precise scientific analysis in collider-based particle physics is possible because of complex simulations that connect fundamental theories to observable quantities.   The significant computational cost of these programs limits the scope, precision, and accuracy of Standard Model measurements and searches for new phenomena.   
We therefore introduce \textit{Deep neural networks using Classification for Tuning and Reweighting} (\DCTR), a neural network-based approach to reweight and fit simulations using all kinematic and flavor information -- the full phase space.  \DCTR can perform tasks that are currently not possible with existing methods, such as estimating non-perturbative fragmentation uncertainties.   The core idea behind the new approach is to exploit powerful high-dimensional classifiers to reweight phase space as well as to identify the best parameters for describing data.  Numerical examples from $e^+e^-\rightarrow\text{jets}$ demonstrate the fidelity of these methods for simulation parameters that have a big and broad impact on phase space as well as those that have a minimal and/or localized impact.  The high fidelity of the full phase-space reweighting enables a new paradigm for simulations, parameter tuning, and model systematic uncertainties across particle physics and possibly beyond. 

\end{abstract}

\maketitle
%% Introduction

In collider-based high-energy physics, parton-, particle-, and detector-level Monte Carlo (MC) simulation programs enable scientific inference by connecting fundamental theories to observable quantities.   However, these tools are often computationally slow and emulate probability distributions that are analytically intractable.  This has resulted in three key simulation challenges for particle physics: (1) an insufficient number of simulated events, (2) unaccounted for biases from simulation parameters, and (3) the inability to utilize the full phase space for parameter tuning.  

%Hoeche:2019rti

A variety of approaches have been proposed to address the above challenges.  The two existing solutions to (1) are to use more~\cite{inproceedings,Farrell:2016ovs,Childers:2015tyv} and/or faster computers or accelerators~\cite{Seiskari:2012uz,Canal:2014gga} or to build fast surrogate models (`fast simulation').  Machine learning tools hold great promise for augmenting~\cite{ATL-SOFT-PUB-2018-002} or replacing~\cite{Paganini:2017dwg,deOliveira:2017pjk,Paganini:2017hrr,deOliveira:2017rwa,Erdmann:2018kuh,Musella:2018rdi,Erdmann:2018jxd,Carminati:2018khv,Chekalina:2018hxi,Hashemi:2019fkn,DiSipio:2019imz,Aaij:2017awb,Monk:2018zsb,ATL-SOFT-PUB-2018-001,Chekalina:2018hxi} current fast detector simulation approaches, but are not yet precise enough to match the full, physics-based detector simulators that are often the limiting factor in the overall software pipeline. Deep learning methods to circumvent expensive simulations for hypothesis testing were studied in the context of effective field theory fits~\cite{Brehmer:2018kdj,Brehmer:2018eca,Brehmer:2018hga}; related ideas will be useful also for reweighting.  The only solution for (2) aside from generating a large set of simulations or interpolating between bins of low-dimensional histograms~\cite{Buckley2009} is to assign event weights for parameter variations.  Currently, this is only possible for a small number of perturbative parameters in parton shower programs~\cite{Mrenna:2016sih,Bellm:2016voq,Bothmann:2016nao} and for parton distribution functions~\cite{Butterworth:2015oua,Buckley:2014ana}.  Pseudo-automated procedures exist for tuning parton shower models~\cite{Buckley2009,Ilten_2017}, but the format of the existing public data means that these algorithms are restricted to a set of mostly one-dimensional inputs that must be assumed to be independent.  The variational method proposed in Ref.~\cite{pmlr-v89-louppe19a} has been demonstrated with high-dimensional data, but utilizes a minimax optimization technique and requires running the simulator many times during training.

This letter introduces \textit{Deep neural networks using Classification for Tuning and Reweighting} (\DCTR, pronounced ``doctor"), a new approach to solve all three computational challenges.  In particular, deep neural network-based classifiers are used to (continuously) reweight one particle-level simulation into another and additionally use the full phase space to fit parameters within a given model.   When the nominal particle-level sample has a corresponding detector-level simulation, then this procedure produces a new detector-level sample as well.  Non-deep machine learning tools have been used in the past for discrete re-weighting~\cite{Martschei:2012pr,Rogozhnikov:2016bdp,Aaij:2017awb,Aaboud:2018htj} with a small number of observables.  Deep-learning-based discrete weighting was considered in~\cite{Aaij:2017awb,Andreassen:2018apy} and continuous single observable reweightings were presented in~\cite{Bothmann:2018trh}.  The re-weighting presented here combines a full-phase space deep learning architecture~\cite{Komiske:2018cqr} with parameterization~\cite{Cranmer:2015bka,Baldi:2016fzo} to fully morph one simulation into another.   There are no restrictions on the size of the input feature space nor on the number of interpolated parameters.  
In addition to re-weighting, we show how \DCTR can be used with a differentiable re-weighting function (such as the one just mentioned) to optimize simulation parameters.  Fitting parameters based on the parameterized classifiers was proposed in Ref.~\cite{Cranmer:2015bka}; here, the fitting procedure uses a classifier to construct the loss function, which can readily incorporate all of the information from the (potentially high-dimensional) input features and be optimized using standard deep learning tools.

The first ingredient to the full phase-space reweighting procedure is a prescription to derive event weights.  Consider two simulations that describe the same phase space $\Omega$ and are described by probability densities $p_0(x)$ and $p_1(x)$, for $x\in\Omega$.  Assuming that $p_0$ and $p_1$ have the same support\footnote{In most physical applications, this is always the case.  If there are regions where $p_0(x)/p_1(x)$ is far from unity, one can add a regularization parameter to the training to mitigate large weights, which may significantly reduce the statistical power of the reweighted dataset.  We found that this works well, but was unnecessary for the examples presented in this paper.}, the function $w(x)=p_0(x)/p_1(x)$ is the ideal per-event weight to morph the second simulation into the first one.   A key observation made by multiple groups in the past is that $w$ can be well-approximated by training a machine learning classifier to distinguish the two simulations.  For example, let $f(x)$ be a neural network and trained with the binary cross-entropy loss:

\begin{align}
\text{loss}(f(x))=-\sum_{i\in \textbf{0}} \log f(x_i)-\sum_{i\in \textbf{1}} \log (1-f(x_i)),
\end{align}

\noindent where $\textbf{0}$ and $\textbf{1}$ represent sets of examples from the two simulations.  Then a well-known result is that\footnote{See Appendix~\ref{sec:derivation} for the derivation.}, $f(x)/(1-f(x))\approx p_0(x)/p_1(x)$.  The benefit of parameterizing $f$ as a neural network is that deep learning can readily analyze all of $\Omega$, which was not possible with shallow learning attempts with a similar statistical foundation.  The closest attempt to a full phase space approach directly tried to learn $p_i(x)$ using the full kinematic (i.e. non-flavor) part of $\Omega$~\cite{Andreassen:2018apy, Andreassen:2019txo}, but this is much harder than learning the ratio. 

An important reweighting scenario is when the two simulations are from the same simulation program, but with different model parameters, $\theta$.  For example, when model uncertainties are evaluated, one may want to transform $p_\theta(x)$ into $p_{\theta+\delta_\theta}(x)$.  When these uncertainties are profiled in a fit, it is important that the transformation procedure be able to continuously interpolate between model parameters.   The neural network reweighting approximation can be extended to this continuous case by adding $\theta$ as a feature~\cite{Cranmer:2015bka,Baldi:2016fzo}: $f(x,\theta)$.  In the examples presented below, the training data are generated with a uniform distribution in $\theta$, but this probability density can be optimized per application and can even be discrete.

Even though generators have many parameters that must be fit to data, gradient methods cannot be used directly with the models as the phase space they produce is not usually differentiable (or at least the derivative is intractable) with respect to their model parameters.  Surrogate generative models built from neural networks can be used for gradient-based parameter fitting, but may not have sufficient quality to be reliable.  Reweighting is a robust alternative to surrogate generative models.  A neural network-based continuous reweighting function is essentially a differentiable (in model parameters) version of the original simulator and can be used to perform inference on the parameters themselves.  This is especially powerful for particle-level parameter tuning to data where one sample with a computational expensive full detector simulation can be continuously reweighted to other parameter points with the same detector model at no extra simulation cost.  

An ideal loss function used to fit model parameters makes use of the full observable phase space.  Typical metrics such as the $\chi^2$ between histogram approximations to probability densities become impractical when $\Omega$ is high dimensional.   As described above, classifiers are powerful tools for accessing all of the available information.  Therefore, one can use a classifier \textit{for the loss}.  When a classifier trained to distinguish some $\boldsymbol{\theta_0}$ from a $\boldsymbol{\theta_1}$ performs poorly, then the two samples are close.  While using classification to quantify differences between event samples has been used for anomaly detection~\cite{DAgnolo:2018cun,Collins:2019jip,Collins:2018epr}, we are unaware of an example where it is used for parameter fitting.   The idea of using the classifier loss as a metric is similar to the minimax strategy in Generative Adversarial Networks~\cite{Goodfellow:2014:GAN:2969033.2969125}, only in this context the generative part is a reweighter and is trained independently.  A more elegant way of implementing this approach is to fit unknown parameters to the values that minimize the nominal classifier loss. In particular, suppose that a reweighter neural network $f$ is trained as described above.  
Such a function will satisfy 
\eqn{f(x,\theta)&=\argmax\limits_{f'}\sum_{i\in \boldsymbol{\theta_0}} \log f'(x_i,\theta)+\sum_{i\in \boldsymbol{\theta}}\log (1-f'(x_i,\theta))
\label{eq:fitpart1}}
for all $\theta$. Note that the $f'$ in the first sum takes the parameter $\theta$ and not $\theta_0$, otherwise the discrimination task would be trivial. 
Now, suppose there is a new sample $\boldsymbol{\theta_1}$ where $\theta_1$ is unknown (for instance, $\boldsymbol{\theta_1}$ are collider data).  
The claim is that if $\theta^*$ is chosen as 
\eqn{\label{eq:fitpart2}
\theta^*&=\argmax\limits_{\theta'}\sum_{i\in \boldsymbol{\theta_0}} \log f(x_i,\theta')+\sum_{i\in \boldsymbol{\theta_1}}\log (1-f(x_i,\theta'))}
then $\theta^*=\theta_1$.  
As $f$ minimizes the cross-entropy loss for any $\theta$ (Eq.~\ref{eq:fitpart1}), 
\eqn{\nonumber\label{eq:fitpart3}
&\sum_{i\in \boldsymbol{\theta_0}} \log f(x_i,\theta_1)+\sum_{i\in \boldsymbol{\theta_1}}\log (1-f(x_i,\theta_1)) \\
& \hspace{3mm}\geq \sum_{i\in \boldsymbol{\theta_0}} \log f(x_i,\theta^*)+\sum_{i\in \boldsymbol{\theta_1}}\log (1-f(x_i,\theta^*))}
must hold.  However, the converse must also be true since $\theta^*$ minimizes the cross-entropy loss as well and therefore, $\theta^*=\theta_1$.  Since $f$ is differentiable, Eq.~\ref{eq:fitpart2} can be solved using standard gradient-based methods.  While Eq.~\ref{eq:fitpart2} performs the fit on the same particle-level phase space as the reweighting, it can be readily extended to do the fitting (via the classification loss) at detector-level while the reweighting can be performed at particle-level using one fully detector-simulated event sample (see Appendix~\ref{sec:alternativefit}).

The last ingredient to \DCTR is a suitable neural network architecture that can effectively capture all the salient features of $\Omega$.  A natural tool for this task is the Particle Flow Network (PFN)~\cite{Komiske:2018cqr}, built on the Deep Sets framework~\cite{DBLP:journals/corr/ZaheerKRPSS17}.  While many deep learning architectures incorporate the symmetries and structure of high energy physics events~\cite{deOliveira:2015xxd,Baldi:2016fql,Barnard:2016qma,Komiske:2016rsd,deOliveira:2017pjk,Louppe:2017ipp,Andreassen:2018apy,Qu:2019gqs,Butter:2017cot,Guest:2016iqz,ATL-PHYS-PUB-2017-003,CMS-DP-2017-049,Sirunyan:2017ezt}, PFNs are particularly effective because they can operate on variable-length sets of particles and respect the quantum-mechanically induced permutation invariance of particle labels.  These networks can also readily incorporate non-kinematic information such as particle flavor.  A particle flow network is a composition of two neural networks $F$ and $\Phi$: $f(\{p_i\})=F(\sum_{i=1}^n\Phi (p_i))$, where $p_i$ is the set of features belong to particle $i$ (momentum and flavor) as well as $\theta$. The function $\Phi$ embeds the input particles into an $\ell$-dimensional \textit{latent space} and $F$ is a simple $\mathbb{R}^\ell\mapsto\mathbb{R}$ neural network.  References~\cite{Komiske:2018cqr,DBLP:journals/corr/ZaheerKRPSS17} proved that this structure is sufficiently flexible to approximate any function and in practice, $\ell\sim\mathcal{O}(10)$.

To illustrate the potential of \DCTR, full phase-space reweighting and parameter tuning is performed on a sample of generated events from the \Pythia 8.230~\cite{Sjostrand:2014zea,Sjostrand:2006za} event generator.  Particle-level $e^+e^-\rightarrow Z\rightarrow \text{dijet}$ events with about 100 particles in each event are clustered into jets using the anti-$k_t$ clustering algorithm~\cite{Cacciari:2008gp} ($R=0.8$) with Fastjet 3.0.3~\cite{Cacciari:2011ma,Cacciari:2005hq}.  The jets are presented to the neural network for training, with each jet constituent represented by $(p_T,\eta,\phi,\text{particle type}, \theta)$, where $\theta$ is the parameter in \Eq{fitpart1}. One million events were generated for each set of \Pythia parameters.  In addition to a default parameter set using the Monash tune~\cite{Skands:2014pea}, three separate samples were generated with uniformly sampled $\alphas$, $\lund$ and $\strange$ in the ranges $[0.10,0.18]$, $[0.50,0.90]$ and $[0.10,0.30]$, respectively.  We also generated one sample where all three parameters were simultaneously uniformly sampled.  These parameters were chosen because they represent both perturbative and non-perturbative physical effects and the ranges are similar to those studied in Ref.~\cite{Ilten_2017}.  The Monash values of the three parameters are 0.1365, 0.68, and 0.217, respectively.

The reweighting and fitting was found to work well without any hyperparameter modifications from Ref.~\cite{Komiske:2018cqr}.  In particular, $\Phi$ has two hidden layers with $\ell=128$ and $F$ is composed of three hidden layers and two output nodes for binary classification, and all the hidden layers have 100 nodes. The activation function used for all layers is ReLu with the exception of the classification output which uses softmax.   All models were implemented in Keras~\cite{keras} with the Tensorflow backend~\cite{tensorflow} and trained using the crossentropy loss with the Adam~\cite{adam} optimizer for 50 epochs, using early stopping with patience 10, with batch size 1000. Each training set contained $8\cdot 10^5$ training and $10^5$ validation jets of each class. Training time was 10-15 min for each model (20 seconds per epoch) on an NVIDIA GeForce GTX 1080.

\begin{figure}[h!]
  \includegraphics[width=1\linewidth]{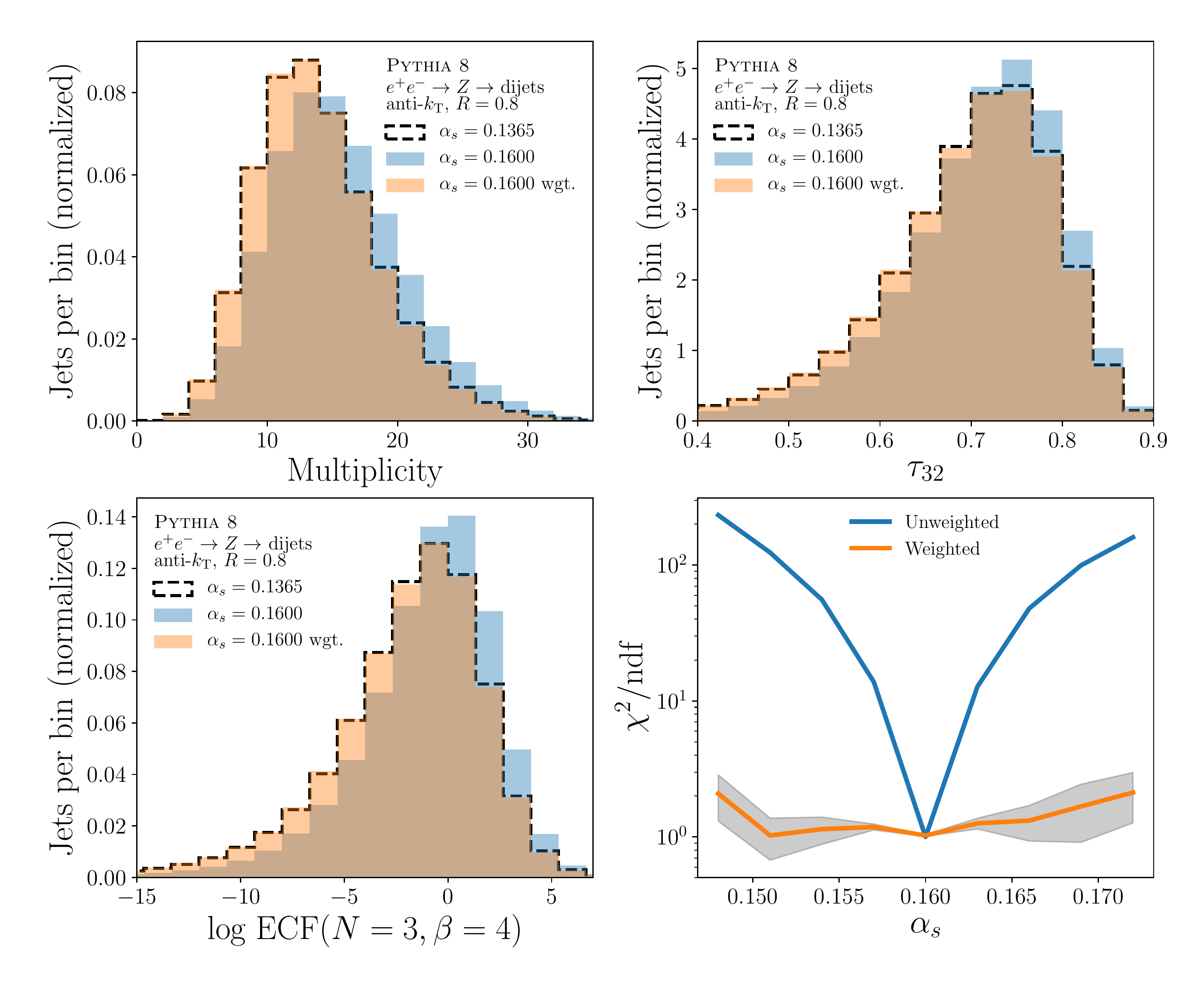} 
\caption{The three histograms shows the result before and after reweighting between two values of $\alphas=\alpha_s$ on different 1D observables. To quantify the quality of the reweighing, and to illustrate one trained model can continuously reweight for any parameter, we show the $\chi^2/\text{ndf}$ for multiplicity as a function of $\alpha_s$ in the lower right plot for reweighting to $\alpha_s=0.1600$. For each value, we compare the $\chi^2$ relative to $\alpha_s=0.1600$ before and after reweighting. Each $\chi^2$ value is averaged over 10 runs and the grey band marks the standard deviation, which is consistent with $\chi^2/\text{ndf}\approx 1$.}
\label{fig:fig1}
\end{figure}

As a first test of \DCTR, a single parameter ($\alphas$) is reweighted using the full phase space of the generated jets.  Results for discrete and continuous reweighting from a varied parameter to the nominal sample are presented in Fig.~\ref{fig:fig1}.  The entire phase-space is reweighted, but is too high dimensional to visualize.  Instead, three histograms of physically relevant one-dimensional observables are presented: the number of particles inside the jet (multiplicity), an $n$-subjettiness ratio $\tau_{32}$~\cite{Thaler:2011gf,Thaler:2010tr}, and a four-point Energy Correlation Function \cite{Larkoski:2013eya} $\text{ECF}(\text{N}=4, \beta=4)$.  By definition, $\tau_{32}=\tau_3/\tau_2$ where $\tau_n=\sum_{i\in\text{jet}}p_{T,i}\min_{j=1...n}\{\Delta R(i,j)\}$ for axis $j$; likewise, $\text{ECF}(\text{N}, \beta)$ is the sum over all quadruples inside the jet weighted by the product of the momenta and the product of all opening angles raised to the power $\beta$.  The large values of $n$, $N$, and $\beta$ are used to expose complex features with a nontrivial dependence on all particles inside the jet. Many more observables were studied, but these ones are representative.

The reweighted distributions are in excellent agreement with the target nominal distribution.  Samples used to make the histograms shown for $\alphas$ values $0.1365$ and $0.1600$ were not used during training or validation.  The fidelity of a continuous reweighting is quantified in the lower right plot of Fig.~\ref{fig:fig1}, which presents the $\chi^2/\ndf$ as a function of the initial $\alpha_s$ parameter value. 

\begin{figure}[h!]
  \includegraphics[width=1\linewidth]{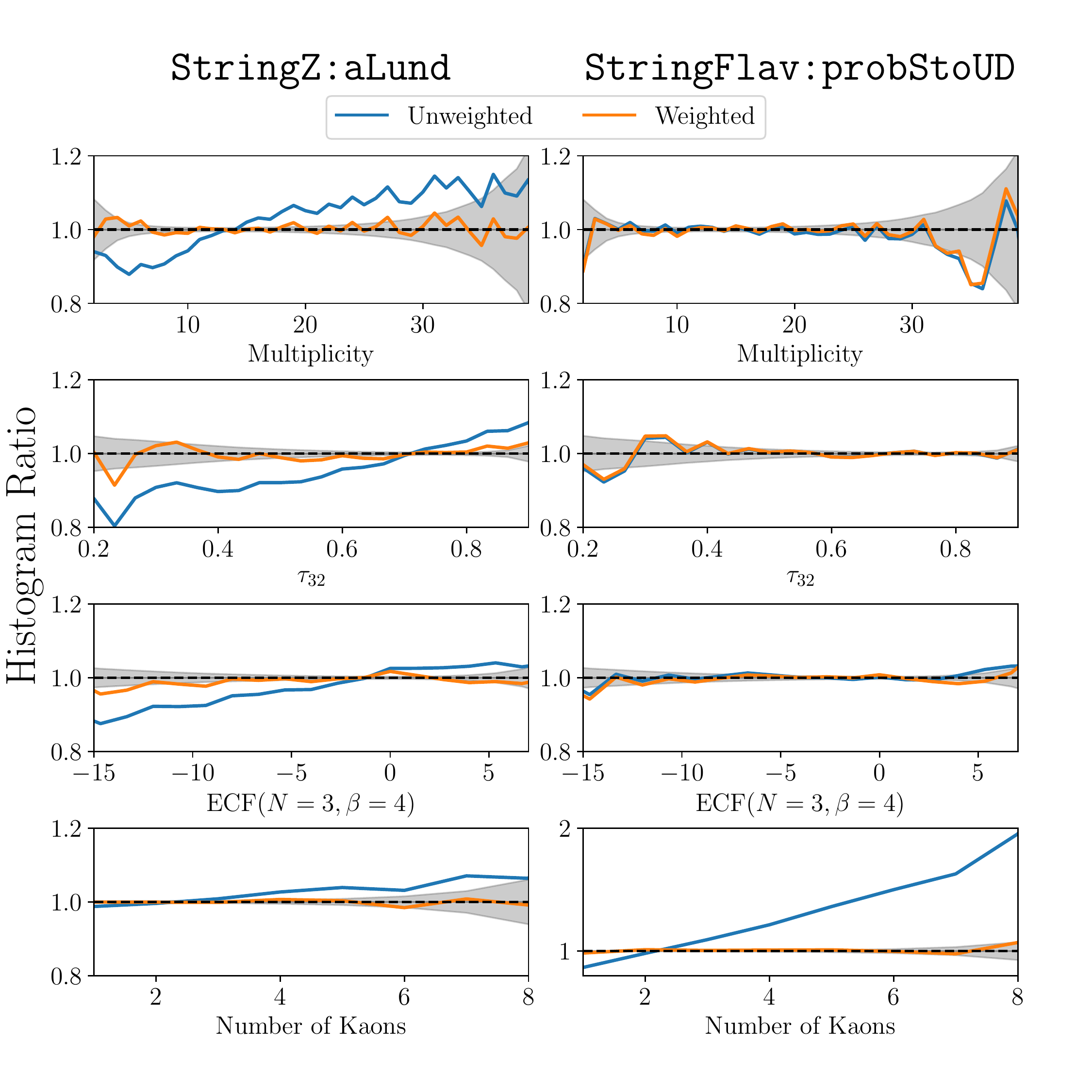} 
\caption{Ratio of histograms from nominal distribution to sample generated with $\lund=0.8$ on the left and $\strange=0.275$ on the right. Both unweighted and weighted histograms ratios are shown. The gray band indicates the statistical uncertainty from both the nominal and variation sample. After reweighting, the ratio only differs from 1 within the statistical uncertainty.}
\label{fig:fig2}
\end{figure}

Variations in $\alphas$ modify many aspects of jet fragmentation and therefore it may be an easy parameter for the reweighting network to learn.  In contrast, the hadronization parameters $\lund$ and $\strange$ may be more difficult because the size of their effects on the phase space is small and/or localized.  Figure~\ref{fig:fig2} shows that despite these potential challenges, the reweighting procedure is able to effectively capture subtle and isolated modifications to the phase space.  Variations in the $\lund$ parameter result in mostly percent-level differences in the presented distributions, which are corrected in the reweighted model.  Modifying the $\strange$ parameter only changes strange particles such as kaons, which highlights the importance of including flavor in the full phase space network.  Importantly, this model learns to only change the distributions related to strange particles, leaving other observables untouched.  Simultaneously reweighting all three parameters also works well, but is more difficult to visualize. We also verified that \DCTR works for $pp$ MC simulations by reweighting from \Pythia to \Herwig for both the quark and gluon samples taken from \cite{Komiske:2018cqr, pathak_aditya_2019_3066475, komiske_patrick_2019_3164691}.

The well-trained \DCTR model can now be used to demonstrate the potential for parameter tuning following Eq.~\ref{eq:fitpart2}.  As a first step, Table~\ref{tab:1dfit} presents the result of a fit where the `data' are the same as the nominal, but with each parameter changed one at a time (each row is a separate fit).  To illustrate the sensitivity to the randomness in the model initialization, each fit is performed ten times.  This variation could be reduced with a more sophisticated neural network and/or more training data.  For each of these one-dimensional fits, the fitted value is consistent with the target value within these statistical fluctuations from initialization, which are $1-3\%$.  As $\alphas$ has a bigger impact on the phase space, it is less sensitive to the initialization statistical fluctuations.  For a fit with data, the statistical and systematic uncertainty could be determined with toys and even profiled, as is standard for parameter fitting.

\begin{table}[h!]
\begin{center}
\caption{Independent fit for simulation where one parameter was changed at a time. The reported numbers are the mean and standard deviation over 10 runs with different model initializations.}
\begin{tabular}{ |c|c|c| } 
 \hline
  \textbf{Parameter} & \textbf{Target value} & \textbf{Fit value} \\ 
  \hline
 \alphas  & 0.1600 & $0.1601\pm 0.0018$ \\ 
 \lund    & 0.8000  &  $0.7980\pm0.0257$\\ 
 \strange & 0.2750 &  $0.2754\pm0.0065$\\ 
 \hline
\end{tabular}
\label{tab:1dfit}
\end{center}
\end{table}

\begin{figure}[h!]
  \includegraphics[width=1\linewidth]{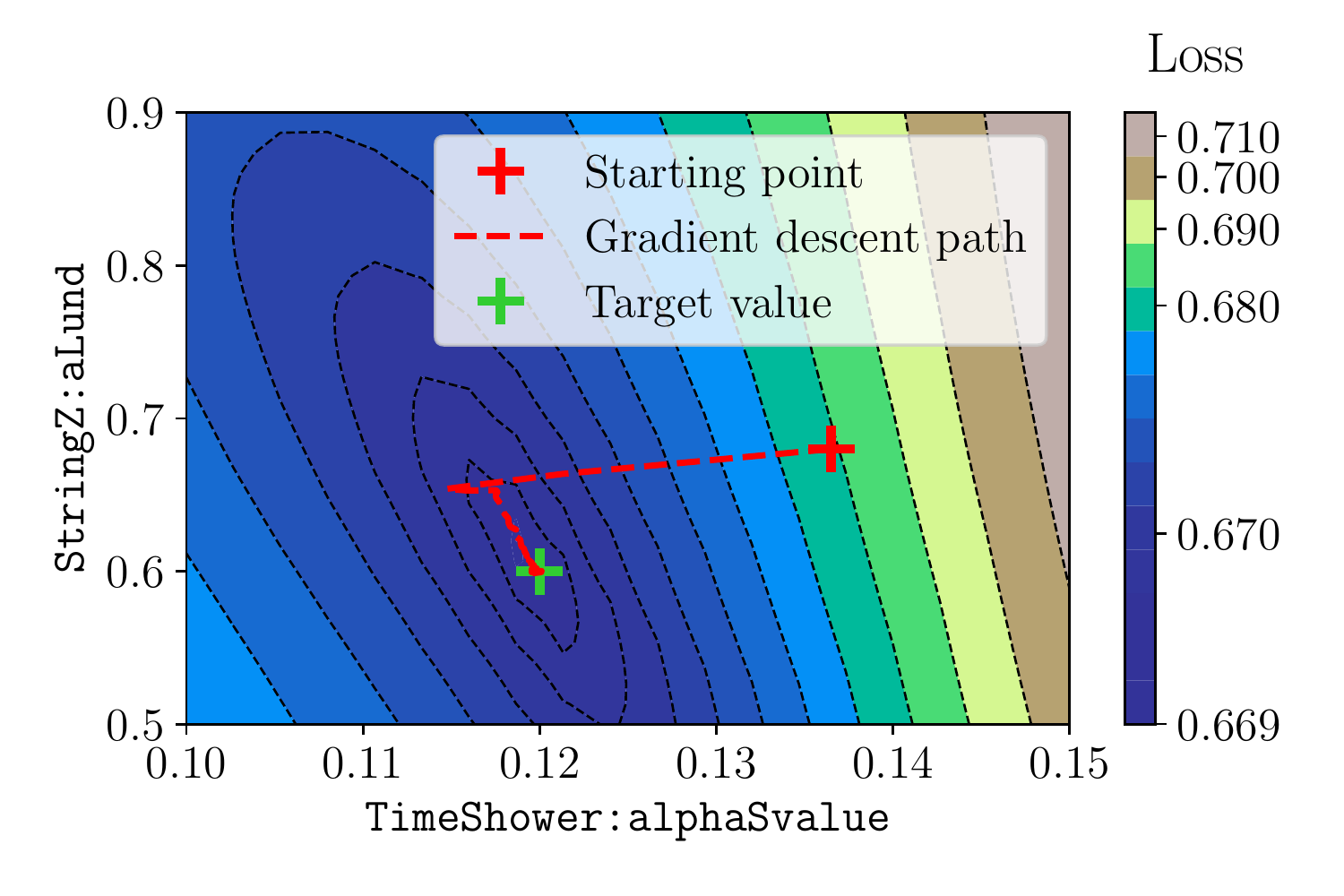}
\caption{Two-dimensional slice through the loss surface for the fit described in \Tab{3dfit}. Markers indicate the starting point at nominal values, the gradient descent path and the target values. From the starting point, gradient descent using Adam overshoots the minimum in its first two epochs before it converges to the target value.}
\label{fig:fig3}
\end{figure}

As a next step, the top part of Table~\ref{tab:3dfit} shows the result of a simultaneous fit to the three parameters.  As with the one-dimensional fit, the fitted values are all statistically consistent with the target values.  Interestingly, the sensitivity to the initialization statistical fluctuations is about the same for the three-dimensional fit as for the one-dimensional fits, providing confidence in the scaling to more parameters.  In practice, the fitting procedure would be validated on a variety of simulations with known parameters, as just described.  
An illustration of the fit itself is shown in Fig.~\ref{fig:fig3}, where a two-dimensional slice through the likelihood landscape is presented and the fit execution demonstrated with markers and dashed lines.  The broadness of the loss in the $\lund$ direction relative to the $\alphas$ one is a reflection of the significantly smaller impact of fragmentation function variations on the observable phase space compared with modifications to the final state shower strong coupling. 
After the validation, the model can be deployed on data, where the parameters are unknown.  The lower part of Table~\ref{tab:3dfit} replicates this scenario, where the Pythia parameters were blinded during the fit.  This closure test indicates that the method is robust to user-bias.

% 3D Fits
\begin{table}[h!]
\begin{center}
\caption{Simultaneous fit for three parameters. The top row shows the results for the validation fit where we knew the target parameters, and the bottom row is the blinded fit. The reported numbers are the mean and standard deviation over 20 runs with different model initializations.}
\begin{tabular}{c|c|c|c| }
 \cline{2-4}
  \multirow{1}{*}{}&\textbf{Parameter} & \textbf{Target value} & \textbf{Fit value} \\ 
  \hhline{-===}
 \multirow{3}{*}{\rotatebox{90}{\textbf{Val.}}}&\alphas  & 0.1200 & $0.1195\pm 0.0022$ \\ 
 &\lund    & 0.6000  &  $0.6276\pm0.0373$\\ 
 &\strange & 0.1200 &  $0.1203\pm0.0071$\\ 
 \hline
 \hline
 \multirow{3}{*}{\rotatebox{90}{\textbf{Blinded}}}&\alphas  & 0.1700 &  $ 0.1707\pm 0.0022 $ \\ 
 &\lund    & 0.7500 &  $ 0.7425\pm 0.0453$\\ 
 &\strange & 0.1400 &  $ 0.1422\pm 0.0065$\\ 
 \hline
\end{tabular}
\label{tab:3dfit}
\end{center}
\end{table}

The empirical results demonstrate that \DCTR is ready to be deployed.  The discrete reweighting could be used to generate new full-detector simulated samples with a different particle-level simulation when at least one fully simulated sample exists.  This could be particularly useful for systematic uncertainties computed using pairs of simulations (e.g. comparing Pythia and Herwig) and for legacy data analysis in which the original detector simulation is no longer available~\cite{Badea:2019vey}.  Continuous reweighting will enable systematic parameter variations for uncertainty estimation that were not possible before (most parameters).  Such variations can even be profiled during any statistical test that fits phase space regions sensitive to the varied nuisance parameters.   Finally, the full power of \DCTR can be used for parameter tuning.  Unlike traditional tuning which use unfolded data that are usually one-dimensional and without observable-observable correlations, a new paradigm is now possible were high-dimensional detector-level data can be used directly.  The full power of the data can be utilized and all of the correlations are correctly accounted for in the fit.  For the first time, this may allow for proper covariance matrices (and thus correlated uncertainties) to be determined for simulation parameter values.  All of these opportunities illustrate the broad applicability of full phase-space reweighting and parameter tuning and the power \DCTR to extend the scope, precision, and accuracy of collider-based particle physics analyses. 

BN would like to thank Luke de Oliveira, Michela Paganini, Chase Shimmin, and Paul Tipton for collaboration on an early stage prototype of this project.  We thank Steve Mrenna and Peter Skands for helpful discussions about automated variations in Pythia.  We also thank Kyle Cranmer, Aviv Cukierman, Patrick Komiske, and Eric Metodiev, and Jesse Thaler for lively discussions about deep learning-based reweighting.  Additionally, we are grateful to Kyle Cranmer, Phil Ilten, and Jesse Thaler for feedback on the manuscript.  This work was supported by the U.S.~Department of Energy, Office of Science under contract DE-AC02-05CH11231. Finally, we are grateful for the opportunity to use the Cori supercomputing resources at NERSC.

\bibliography{myrefs}

\appendix

\section{Optimal Functions}
\label{sec:derivation}

The results presented here can be found (as exercises) in textbooks, but are repeated here for easy access.   Let $X$ be some discriminating features and $Y\in\{0,1\}$ is another random variable representing class membership.  Consider the general problem of minimizing some average loss for the function $f(x)$:

\begin{align}
\label{eq:loss}
f=\text{argmin}_{f'}\mathbb{E}[\text{loss($f'(X),Y$)}],
\end{align}

\noindent where $\mathbb{E}$ means `expected value', i.e. average value or mean (sometimes represented as $\langle \cdot\rangle$).  The expectation values are performed over the joint probability density of $(X,Y)$.  One can rewrite Eq.~\ref{eq:loss} as

\begin{align}
\label{eq:loss2}
f=\text{argmin}_{f'}\mathbb{E}[\mathbb{E}[\text{loss($f'(X),Y$)}|X]].
\end{align}

\noindent The advantage\footnote{The derivation below for the mean-squared error was partially inspired by Appendix A in Ref.~\cite{Cranmer:2015bka}.} of writing the loss as in Eq.~\ref{eq:loss2} is that one can see that it is sufficient to minimize the function (and not functional) $\mathbb{E}[\text{loss($f'(x),Y$)}|X=x]$ for all $x$.  To see this, let $g(x)=\text{argmin}_{f'}\mathbb{E}[\text{loss($f'(x),Y$)}|X=x]$ and suppose that $h(x)$ is a function with a strictly smaller loss in Eq.~\ref{eq:loss} than $g$.  Since the average loss for $h$ is below that of $g$, by the intermediate value theorem, there must be an $x$ for which the average loss for $h$ is below that of $g$, contradicting the construction of $g$.

Now, consider the case where the loss is cross-entropy:

\begin{align}
\label{eq:loss3}
&\max_z\mathbb{E}[Y\log(z)+(1-Y)\log(1-z)|X]\\
\hspace{2mm}&=\max_z\left(\mathbb{E}[Y|X]\log(z)+(1-\mathbb{E}[Y|X])\log(1-z)\right),
\end{align}

\noindent where $z=f'(x)$ is fixed.  Equation~\ref{eq:loss3} is maximized for $g(x)=\mathbb{E}[Y|X=x]$.  Coincidentally, the exact same result holds if using mean squared error loss. When using either loss function with two outputs and the softmax activation for the last neural network layer, the first output will asymptotically approach $g(x)$ and the other by construction will be $1-g(x)$.  The ratio of these two outputs is then:

\begin{align}
\frac{g(x)}{1-g(x)}&=\frac{\mathbb{E}[Y|X=x]}{\mathbb{E}[1-Y|X=x]}\\
&=\frac{\Pr(Y=1|X=x)}{\Pr(Y=0|X=x)}\\
&=\frac{p(X|Y=1)\Pr(Y=1)}{p(X|Y=0)\Pr(Y=0)}\\
&=\text{Likelihood ratio}\times\frac{\Pr(Y=1)}{\Pr(Y=0)}.
\end{align}

\noindent Therefore, the output is proportional to the likelihood ratio.  The proportionality constant is the ratio of fractions of the two classes used during the training.  In the paper, the two classes always have the same number of examples and thus this factor is unity.

\section{Alternative Fitting Method}
\label{sec:alternativefit}

In the main body, it was shown how a continuously parameterized NN used for reweighting:

\begin{align}
f(x,\theta)=\argmax\limits_{f'}\sum_{i\in\bf{\theta_0}}\log(f'(x_i,\theta))+\sum_{i\in\bf{\theta}}\log(1-f'(x_i,\theta))
\end{align}

\noindent can also be used for fitting:

\begin{align}
\theta^*=\argmax\limits_{\theta'}\sum_{i\in\bf{\theta_0}}\log(f'(x_i,\theta'))+\sum_{i\in\bf{\theta}}\log(1-f'(x_i,\theta')).
\end{align}

\noindent This works well when the reweighting and fitting happen on the same `level'. However, if the reweighting happens at truth level (before detector simulation) while the fit happens in data (after the effects of the detector), this procedure will not work. It works only if the reweighting and fitting both happen at detector-level or both happen at truth-level.  The following is an alternative method:

\begin{align}\nonumber
\theta^*&=\argmax\limits_{\theta'}\min\limits_{g}\sum_{i\in\bf{\theta_0}}\log(g(x_i))\\
&\hspace{6mm}+\sum_{i\in\bf{\theta}}w(x_i,\theta)\log(1-g(x_i)),
\end{align}

\noindent where $w(x_i,\theta)=f(x_i,\theta)/(1-f(x_i,\theta))$ is a trained \DCTR using binary cross entropy as in the main body.   The intuition of the above equation is that the classifier $g$ is trying to distinguish the two samples and we try to find a $\theta$ that makes $g$'s task maximally hard.  If $g$ cannot tell apart the two samples, then the reweighting has worked.  This is similar to the minimax graining of a GAN, only now the analog of the generator network is the reweighting network which is fixed and thus the only trainable parameters are the $\theta'$.  The advantage of this second approach is that it readily generalizes to the case where the reweighting happens on a different level:

\begin{align}\nonumber
\theta^*&=\argmax\limits_{\theta'}\min\limits_{g}\sum_{i\in\bf{\theta}_0}\log g(x_{D,i})\\
&\hspace{6mm}+\sum_{i\in\bf{\theta}}w(x_{T,i},\theta)\log (1-g(x_{D,i})),
\end{align}

\noindent where $x_T$ is the truth value and $x_D$ is the detector-level value.  In simulation (the second sum), these come in pairs and so one can apply the reweighting on one level and the classification on the other.  

Asymptotically, both this method and the one in the body of the DCTR paper learn the same result: $\theta^*=\theta_0$.   To see this for the second method, consider the same logic as in Appendix~\ref{sec:derivation}.  Conditioning on $x$ and $\theta$, the optimal $g$ is given by

\begin{align}
g=\frac{\mathbb{E}[Y|X=x]}{(1-\mathbb{E}[Y|X=x])w(x,\theta)+\mathbb{E}[Y|X=x]},
\end{align}

\noindent which reduces to the result of the previous appendix when $w=0$.  For fixed $g$, the loss is maximized when $g$ is independent of $x$, which happens if $(1-\mathbb{E}[Y|X=x])w(x,\theta)\propto \mathbb{E}[Y|X=x])$, which means that $w(x,\theta)$ is proportional to the likelihood ratio between the two samples.  An example implementation of this method in Keras can be found at Ref.~\cite{dctrgithub}.

\end{document}